\documentclass{svproc}
\usepackage{amsmath,amsfonts}
\usepackage{algorithmic}
\usepackage{algorithm}
\usepackage{array}
\usepackage[caption=false,font=normalsize,labelfont=sf,textfont=sf]{subfig}
\usepackage{textcomp}
\usepackage{stfloats}
\usepackage{url}
\usepackage{verbatim}
\usepackage{graphicx}
\usepackage{cite}

\usepackage{bm}

\begin{document}
\mainmatter              % start of a contribution
\title{HyperS2V: A Framework for Structural Representation of Nodes in Hyper Networks}
\titlerunning{HyperS2V}  % abbreviated title (for running head)
\author{Shu Liu*\inst{1} \and Cameron Lai\inst{1}  \and Fujio Toriumi\inst{1}  }
\authorrunning{Liu et al.} % abbreviated author list (for running head)
\tocauthor{Shu Liu, Cameron Lai, and Fujio Toriumi}
\institute{The University of Tokyo, 7-3-1 Hongo, Bunkyo-ku, Tokyo, 113-8654, Japan,\\
% \email{\{liu,nishiguchi\}@torilab.net, tori@sys.t.u-tokyo.ac.jp}
\email{shu.liu.eq@gmail.com, cameron.lai@torilab.net, tori@sys.t.u-tokyo.ac.jp}}
% \institute{Princeton University, Princeton NJ 08544, USA,\\
% \email{I.Ekeland@princeton.edu},\\ WWW home page:
% \texttt{http://users/\homedir iekeland/web/welcome.html}
% \and
% Universit\'{e} de Paris-Sud,
% Laboratoire d'Analyse Num\'{e}rique, B\^{a}timent 425,\\
% F-91405 Orsay Cedex, France}

\maketitle              % typeset the title of the contribution

\begin{abstract}
% Unlike ordinary (simple) networks, hyper networks can represent higher-order interactions between nodes and store a wealth of information. This feature has been widely observed in the real world. Learning embedded representations of nodes is a method that maps network structures to lower-order spaces, allowing machine-learning methods for vector data to be applied to network data as well. However, the embedded representation learning method focusing on the structure needs to be explored. In this study, we propose HyperS2V, a node embedding method based on the structural similarity of hyper networks.
% We first define the hyper-degree to retain the structural information of nodes in hyper networks. We then create a novel function to quantify the structural similarity between hyper degrees. Lastly, we generate structural embeddings following the framework of struc2vec.
% We also conducted intrinsic and extrinsic experiments on toy and real networks and showed the superiority of HyperS2V in both interpretability and adaptability to downstream tasks.
In contrast to regular (simple) networks, hyper networks possess the ability to depict more complex relationships among nodes and store extensive information. Such networks are commonly found in real-world applications, such as in social interactions. Learning embedded representations for nodes involves a process that translates network structures into more simplified spaces, thereby enabling the application of machine learning approaches designed for vector data to be extended to network data.
Nevertheless, there remains a need to delve into methods for learning embedded representations that prioritize structural aspects. This research introduces HyperS2V, a node embedding approach that centers on the structural similarity within hyper networks. Initially, we establish the concept of hyper-degrees to capture the structural properties of nodes within hyper networks. Subsequently, a novel function is formulated to measure the structural similarity between different hyper-degree values. Lastly, we generate structural embeddings utilizing a multi-scale random walk framework.
Moreover, a series of experiments, both intrinsic and extrinsic, are performed on both toy and real networks. The results underscore the superior performance of HyperS2V in terms of both interpretability and applicability to downstream tasks.
\end{abstract}
\section{Introduction}
Networks are composed of entities known as nodes, with connecting relationships established through entities called edges \cite{10.1093/acprof:oso/9780199206650.001.0001}. In simple networks, edges represent interactions between pairs of nodes. However, real-world scenarios often involve interactions between more than two nodes \cite{battiston2020networks}. Examples include group discussions in information diffusion networks \cite{wasserman_faust_1994,mastrandrea2015contact}, competitive relationships among multiple species in ecosystem networks \cite{case1981testing} and functional interactions across multiple brain regions in brain networks \cite{doi:10.1073/pnas.1019641108}.
These interactions that involve more than two nodes cannot be adequately captured in simple networks, where they are represented as cliques (i.e. all nodes are connected to each other). This characteristic can lead to a loss of fidelity to reality \cite{10.1145/1401890.1401971}. Hyper networks address this limitation by allowing for interactions among multiple nodes through the use of hyperedges \cite{battiston2020networks}.

Node embedding is the process of assigning nodes from a network to vectors in a lower-dimensional space while preserving their inherent characteristics \cite{goyal2018graph, zhou2022network}. This technique bridges the gap between the network's structure and machine learning methods. Nodes with similar characteristics should be situated close to each other in this vector space. These characteristics fall into two main categories: those related to proximity, which concern the spatial arrangement of nodes within the network, and those linked to structure, which involve the connectivity patterns among nodes \cite{ahmed2020role}.
The latter category measures dissimilarities in nodes' connection patterns. Nodes that share similar degrees with both themselves and their neighbors demonstrate likeness, even if they belong to different components (i.e. non-connected sub-networks within a larger network) \cite{rossi2020proximity}. This characteristic allows for the assessment of nodes across various components or networks. An exciting application therefore lies in transferring node label information between networks using domain adaptation techniques applied to the acquired embeddings \cite{liu2021multiple, liu2022flexible}. 
% Additionally, our preliminary research showcases the exceptional performance of structural embeddings in predicting link signs and directions within signed directed networks \cite{liu2023signeds2v}.
Additionally, previous research has showcased the exceptional performance of structural embeddings in predicting link signs and directions within signed directed networks \cite{liu2023signeds2v}.

Numerous node embedding methods have been devised for simple networks, based on either proximity (e.g., DeepWalk \cite{perozzi2014deepwalk}, node2vec \cite{grover2016node2vec}, LINE \cite{tang2015line}) or structure (e.g., struc2vec \cite{ribeiro2017struc2vec}, GraphWave \cite{donnat2018learning}, SignedS2V \cite{liu2023signeds2v}). However, when it comes to hyper networks, only a handful of proximity-based methods, such as HHE \cite{ZHU2016150} and HGE \cite{10.1145/3269206.3269274}, exist. Notably, despite the considerable expressive potential and practical applications of hyper networks, no structure-based method for node embedding has been developed.

Addressing this gap, we propose HyperS2V, a method for learning structure-based node embeddings in hyper networks. Through a range of experiments conducted on both toy and real networks, we demonstrate the superior interpretability and adaptability of HyperS2V in various downstream tasks. Furthermore, we present a case study illustrating the practical application of structural embeddings derived from hyper networks.
The contributions of this work contain the following:
\begin{enumerate}
    \item Proposing HyperS2V, a novel structure-based embedding method for hyper networks via hyper-degree and distance function.
    \item Demonstrating the outstanding interpretability from toy networks and adaptability to hyperedge prediction and hyperedge dimension prediction on real networks.
    \item Showcasing the usage of structural embeddings on the Les Misérables dataset, enabling the identification of meaningful clusters of characters based on their roles and interactions within the novel, and highlighting the potential for further understanding the network's structural similarities among unrelated roles.
    % , opening up avenues for deeper analysis of narrative connections in literary networks.
\end{enumerate}

\section{Preliminary}
Let $H=(V,E)$ denote a hyper network that contains $|V|$ nodes and $|E|$ hyperedges. $v_i\in V$, $e_j\in E$, $e_j \subseteq V$. If $|e_i|=2$ for all $e_i \in E$, then $H$ degenerates to a simple network. We assume all the hyperedges involve two or more nodes, namely, $|e_i| \ge 2$.
The incidence matrix $I=\{0,1\}^{|V|\times |E|}$ represents the relationship between nodes and hyperedges:
\begin{equation}
    I_{ij}=\left\{
    \begin{alignedat}{1}
    1&\quad if\enspace v_i \in e_j, \\
    0&\quad otherwise.
    \end{alignedat}
    \right.
    \label{formula:Iij}
\end{equation}
The degree of node $v_i$ is $\sum_{j}{I_{ij}}$, and the size of edge $e_j$ is $\sum_i{I_{ij}}$.
Node embedding task for hyper networks aims to find the function $f:V\rightarrow \mathbb{R}^{|V|\times k}$ that maps nodes to $k-$dimensional vectors, where $k\ll |V|$ and $k\ll |E|$.

In complex networks, different types of equivalence are proposed based on the specified equivalence relation \cite{rossi2014role, Jin2022TowardsUA}.
In this work, we focus on automorphic equivalence while considering the potential of preserving graph-theoretic properties and the proper level of strictness. 
Two nodes have automorphic equivalence if they share the same connection patterns: graph-theoretic properties, such as +out, +in, -out, and -in degrees, and centralities\cite{10.2307/270991}.
In this paper, we use structural equivalence to represent automorphic equivalence.

Distinct forms of equivalence are put forth in complex networks, grounded in the specified equivalence relationship \cite{rossi2014role,Jin2022TowardsUA}. 
This study focuses on automorphic equivalence, while taking into account the potential for maintaining graph-theoretical characteristics and appropriate flexibility. 
In simple networks, two nodes exhibit automorphic equivalence if they possess the same connection patterns, which encompass graph-theoretical attributes (e.g. degrees), as well as centralities \cite{10.2307/270991}.
We broaden the concept of automorphic equivalence from simple networks to hyper networks by considering the sizes of the connected hyperedges for a given node.
It is important to recognize that hyperedges can take on various structural configurations, and numerous researchers have investigated methods to assess these structures using measures such as encapsulation \cite{larock2023encapsulation} and downward inclusion \cite{landry2023simpliciality}.
To put it simply, two nodes are considered locally automorphically equivalent if they share both the same number of hyperedges and the sizes of these hyperedges. 
Therefore, automorphic equivalence within hyper networks is more lenient compared to that within basic networks. In this paper, we employ structural similarity as a means to express automorphic equivalence.

\section{Related research}
There are several proximity-based embedding methods proposed for nodes in hyper networks. One such method is HHE \cite{ZHU2016150}, which was proposed for document recommendation tasks where the hyperedges involve different interactions between heterogeneous entities. HHE generates $k-$dimensional embeddings by seeking the first $k$ generalized eigenvectors corresponding to the $k$ smallest nonzero eigenvalues of the normalized objective matrix derived from the Laplacian matrix of the hyper networks. Therefore, HHE is an embedding method based on the Laplacian matrix, which focuses on the nodes' connectivity. Consequently, HHE is a proximity-based embedding method.
Another approach is HGE \cite{10.1145/3269206.3269274}, a deep-learning model that releases the constraint of relations from pairwise to multi-body to fit for hyper networks. HGE generates embeddings by optimizing a function such that nodes within any hyperedges should have similar embeddings but nodes with no hyperedges should be vastly different. Thus, HGE is also a proximity-based embedding method.
DHNE \cite{tu2018structural}, on the other hand, is a deep-learning model proposed for hyper networks with uniformly sized (limited to three) hyperedges. DHNE applies an optimization function that preserves first-order proximity (nodes within any hyperedges should have similar embeddings) and second-order proximity (nodes sharing common neighborhoods should have similar embeddings).
% Even though DHNE takes the neighborhood similarity into embeddings, which is still the closeness of the node location, thus, it belongs to the proximity-based category.
DHNE considers neighborhood similarity in embeddings. This is still related to the closeness of node locations, and is therefore a proximity-based method
Since the size of the hyperedge is limited to three, DHNE cannot be applied to general hyper-network embedding tasks.

There are some works on structure-based node embedding for simple networks. Struc2vec \cite{ribeiro2017struc2vec} is a famous method among them. It collects structural similarities between each node pair in multiple scales, from nodes to multiple hops neighbor's degrees, and builds a multi-layer network where the similarity in a specified scale constructs each layer. Struc2vec learns embeddings by applying Skip-Gram \cite{mikolov2013efficient} on the walk sequence generated through biased random walks on the multi-layer network. Since struc2vec evaluates the structural similarity of all node pairs, two nodes could be placed near each other as long as they have similar connection patterns, even if they are located within two different components. Structural embedding methods are proposed for different types of simple networks (such as signed and directed networks) inspired by struc2vec \cite{liu2023signeds2v}.
GraphWave \cite{donnat2018learning} employs a unique approach by considering wavelets as probability distributions across the graph, focusing on how diffusion spreads rather than its location. The method utilizes the empirical characteristic function to embed these wavelet distributions to create embeddings. 
Role2vec \cite{ahmed2020role} learns the embeddings of roles (structural label of nodes) instead of nodes. It assigns role labels to nodes by structural features and applies the skip-gram model on walk sequences of roles sampled from random walks. The embedding of nodes is the same if they belong to an identical role.

However, based on our current understanding, no methods focus on embedding nodes in hyper networks using structural information. This research endeavor aims to address this gap.

\section{Proposed method}
\begin{figure}[t]
    \centering
    \includegraphics[scale=0.26]{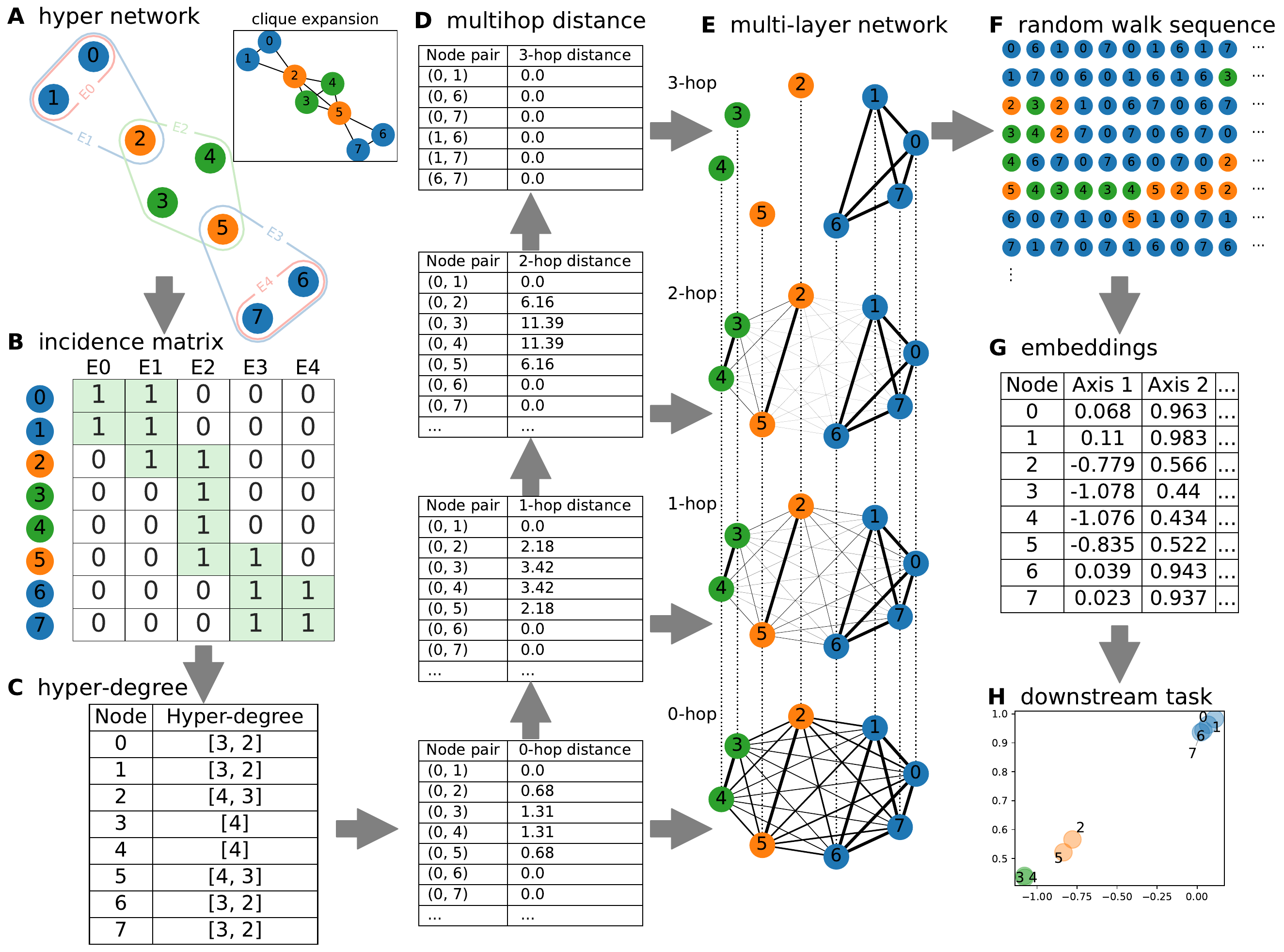}
    \caption{Conceptual sketch of HyperS2V}
    \label{fig:overall_flow}
\end{figure}
Structural information of a node in simple networks derives from the connection pattern of the node, namely, the degree of the node and the surrounding nodes. This principle also works in hyper networks, but two significant problems exist: 1) how to represent the degree information and 2) how to calculate the distance (measure of structural difference) between nodes in a hyper network.
We first define the hyper-degrees (HD) of nodes in a hyper network to represent the degree information and propose a function to calculate the structure-based distance between two hyper-degrees. Then, we apply the random walk approach proposed by struc2vec to generate structural embeddings.
\subsection{Hyper-degree}
The degree of nodes in simple networks is a scaler representing the number of corresponding pair-wise edges. In the hyper network context, an edge could involve more than two nodes; namely, the edge size could be more than two. Thus, it is critical to evaluate not only the number of edges but also the size of each edge while assessing the structure of a node.
Therefore, we define the hyper-degree of node $i$ in hyper network $H$ as follows.
\begin{equation}
    HD_i=sort\bigg(\big\{\sum_{v}{I_{vj}}|I_{ij}=1\big\}\bigg),
\end{equation}
where $I$ indicate the incidence matrix, $\sum_{v}{I_{vj}}$ is the size of edge $e_j$, and $sort(\cdot)$ is a descending-order sort function.
The length of $HD_i$ is the degree of node $i$, and each element in $HD_i$ is the size of the corresponding edge that $i$ involved; we use hyper-degree to combine the degree and edges' sizes to indicate the $0-$hop connection pattern of a node in a hyper network.
\subsection{Magnitude-position distance between hyper-degrees}
We empirically create a novel distance function to assess the distance between two hyper-degrees.
% We propose using the following equation to calculate the structural distance for each pair of elements, a pair of edge degrees.
Given two nodes $u$ and $v$, with the hyper-degrees $HD_u$ and $HD_v$, $s_{u_i}$ and $s_{v_j}$ are $i$th and $j$th elements in $HD_u$ and $HD_v$, representing $i$th and $j$th largest edges' sizes corresponding to $u$ and $v$, respectively.
The magnitude-position distance (MPD) between $s_{u_i}$ and $s_{v_j}$ is calculated by the following equation.
\begin{equation}
    MPD(s_{u_i},s_{v_j})=\exp{\bigg(\sqrt[n]{\big(1-\frac{min(s_{u_i},s_{v_j})}{max(s_{u_i},s_{v_j})}\big)^n+|b_{u_i}-b_{v_j}|^n}\bigg)}-1
    \label{equ:mpd}
\end{equation}
$n \in [1,2,3,...]$ is the exponent to control the magnitude of the values and sensitivity to outliers; here, we set $n$ to 2.
$1-\frac{min(s_{u_i},s_{v_j})}{max(s_{u_i},s_{v_j})}$ is the term of magnitude, and $|b_{u_i}-b_{v_j}|$ is that of positional importance. 
$b_{u_i}$ and $b_{v_j}$ are the bias terms indicating their ``positional importance" in the sequence $HD_u$ and $HD_v$.
For $s$ in descending-order sorted list $HD$,
the bias $b_s$ is calculated by the equation.
\begin{equation}
    b_{s}=\frac{1}{max(HD)-s+1}
    \label{equ_importance}
\end{equation}
The bias of the largest element in HD is 1, indicating that the largest edge is in the most important position in a node's HD. As the element decreases, the less positional importance it holds, and the smaller the bias becomes.
For instance, the figure plots the bias for each degree in a list of $[11, 10, ..., 3, 2]$.

\begin{figure}[t]
    \centering
    \includegraphics[scale=0.52]{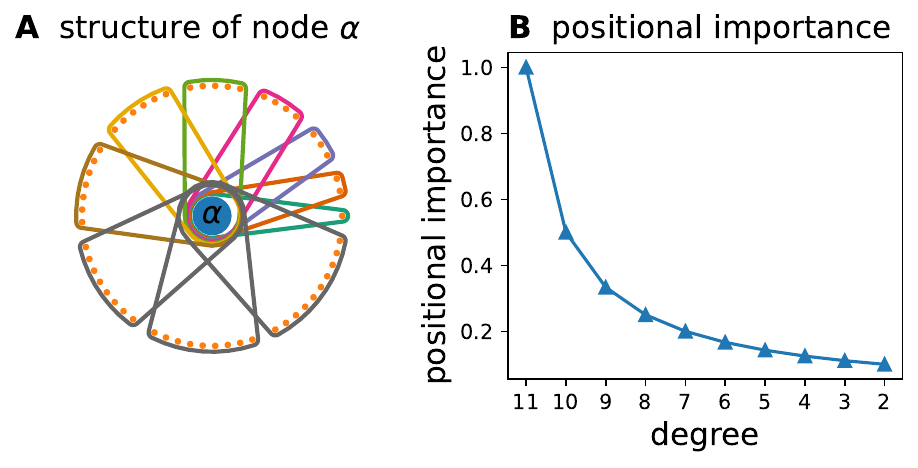}
    \caption{Example of positional importance}
    \label{fig:MPD_CMPD}
\end{figure}

We apply $1-\frac{min(\cdot)}{max(\cdot)}$ to constrain the difference in magnitude within a range of $[0,1)$ to match the range of the bias term, which is also $[0,1)$.
Overall, we assess the distance between two edge sizes in HD by leveraging the difference in magnitude and importance. 
The more difference in $s_{u_i}$ and $s_{v_j}$ and the more difference in positional importance, the larger distance is between $u_i$ and $v_j$.

Since the number of corresponding edges of two nodes could differ, hyper-degrees' lengths vary. We apply dynamic time wrapping (DTW) \cite{salvador2007toward} to calculate the distance of two lists with different lengths. DTW compares the distance between each pair of elements from the two input lists and searches for the path with the lowest summation of distance. The overall distance $D^{'0}$ between $u$ and $v$ can be calculated as the following equation. 
\begin{equation}
    D^{'0}(u,v)=DTW(MPD;\enspace HD_u,\enspace HD_v).
    \label{equ:dis_0_mpd}
\end{equation}
DTW calculates the MPD between all pairs of $(s_{u_i},s_{v_j}) \in HD_u\times HD_v$ and finds the optimal alignment between the two sequences $HD_u$ and $HD_v$ as the final $0-$hop distance.
% \subsection{Sensitivity confirmation of magnitude-position distance}
% We show the sensitivity of magnitude-position distance in two aspects, the largest edge's size and the length of HD.
% We check the distance when the largest edge's size increases while increasing the lengths of HD.
% For $x$ and $y$ are integers ranging from 0 to 9 separately, $HD1 = [3+x+y, 2+y, .., 2]$. The largest element of HD1 is $x$ larger than the second. Compared to $HD2 = [3, 2]$, HD1 is $y$ longer. Thus, we use $\delta max$ to represent $x$ and $\delta len$ for $y$. Note that we eliminate the duplicates in HD1.
% % HD2 = [3,2], and HD1 ranges from [3,2], [4,2], ..., [11,2]; [4,3,2], [5,3,2], ..., [11,3,2]; ...; [12,11,...,3,2], [13,11,...,3,2], ..., [21,11,...,3,2].
% The distance between HD1 and HD2 is shown in the figure.

% [figure]

% The distance increases when the $\delta max$ or $\delta len$ increases, but the increase rates differ: concave to $\delta max$ but convex to $\delta len$. This result indicates that the distance is more sensitive to the length of HD (the degree of the node) than the max size of the corresponding edges. We discuss that this preference is more general and suitable for structural embedding and the potential downstream tasks in hyper networks.
\subsection{Processing the duplicated elements in HD}

% Since the hyperedges' size in hyper networks repeats with considerable frequency, it is reasonable that duplicated elements exist in HD generally. 
As a given hyperedge size will frequently repeat in a hyper network, therefore it is reasonable to assume that duplicated elements exist in the HD. 
We find that the property of MPD is insufficient in the case of the duplicated elements in HD.
To simplify the analysis, we degenerate the network into a simple network; namely, all edges are pairwise. Thus, all the elements in HD are two. Considering two nodes $u$ and $v$ in a simple network, $d_u=2$ and $d_v=3$, (i.e., $HD_u=[2,2]$ and $HD_v=[2,2,2]$ in hyper-degree formation.) 
From equation \ref{equ:mpd}, for any element pair $(s_{u_i}, s_{v_j}) \in HD_u \times HD_v$, $MPD(s_{u_i}, s_{v_j})=0$, because $s_{u_i}=s_{v_j}=2$ and $b_{u_i}=b_{v_j}=1$. Consequently, the distance between $u$ and $v$ will be 0 by equation \ref{equ:dis_0_mpd}. 

To fix this problem, we compress HD by frequency and form the collapsed HD, a list of 2-dimensional tuples where the first element is the edge size and the second is the frequency.
For $HD = [s_1^1, \cdots, s_1^{f_1}, s_2^1, \cdots, s_2^{f_2}, \cdots]$, the collapsed HD $CHD = [(s_1, f_1), (s_2, f_2), \cdots]$.
For instance, $HD_a = [4,4,3,2,2,2]$, then $CHD_a=[(4,2), (3,1), (2,3)]$.
We first calculate the position importance $[b_{s_1}, b_{s_2}, \cdots]$ of $[s_1, s_2, \cdots]$ by equation \ref{equ_importance}. Unlike HD, we divide $b_{s_i}$ by the frequency $f_i$ to simulate that an additional edge of the same size contributes less to the structure for the specified size.
% that the more edges with the same size, the less contribution each edge has on the structure of a specified size.

Given two nodes $u$ and $v$, with the collapsed hyper-degrees $CHD_u$ and $CHD_v$, $(s_{u_i}, f_{u_i})$ and $(s_{v_j}, f_{v_j})$ are $i$th and $j$th tuples in $CHD_u$ and $CHD_v$, representing $i$th and $j$th largest edges' sizes and the corresponding frequency, respectively.
The collapsed magnitude-position distance (CMPD) between $s_{u_i}$ and $s_{v_j}$ is calculated by the following equation.
\begin{equation}
    CMPD(s_{u_i},s_{v_j})=max(f_{u_i},f_{v_j}) \Bigg(\exp{\bigg(\sqrt[n]{\Big(1-\frac{min(s_{u_i},s_{v_j})}{max(s_{u_i},s_{v_j})}\Big)^n+\bigg|\frac{b_{u_i}}{f_{u_i}}-\frac{b_{v_j}}{f_{v_j}}\bigg|^n}\bigg)}-1\Bigg).    
    \label{equ:cmpd}
\end{equation}

Given the previously outlined distance of node $u$ ($HD_u=[2,2]$) and $v$ ($HD_v=[2,2,2]$), now that we involve frequency in the positional importance term, $CMPD(s_{u_i}, s_{v_j})> 0$.
Moreover, grouping the duplicated elements would dramatically reduce the computational cost and potentially lead to a more precise distance assessment.

The following equation calculates the overall $0-$hop distance ($D^0$) between two nodes $u$ and $v$.
\begin{equation}
    D^0(u,v)=DTW(CMPD;\enspace CHD_u,\enspace CHD_v).
    \label{equ:dis_0_cmpd}
\end{equation}
% DTW calculates the CMPD between all pairs of $(s_{u_i},s_{v_j}) \in CHD_u\times CHD_v$ and finds the optimal alignment between the two sequences $CHD_u$ and $CHD_v$ as the final $0-$hop distance.

% \subsection{Sensitivity confirmation of collapsed magnitude-position distance}
% We confirm the sensitivity of CMPD against the frequency and duplicated position.
% Let $HD_1 = [11, 10, ..., 3, 2]$ as the comparison HD, the $CHD_1$ is $[(11, 1), (10, 1), \cdots, (3, 1), (2, 1)]$.
% For $x$ and $y$ are integers ranging from 0 to 9 separately, $CHD_2$ is $[(11, 1), \cdots, (2+y, x), \cdots, (2, 1)]$. $y$ is the position index of the duplicated element, the bigger, the larger duplicated edge's size.
% And $x$ is the frequency of the specified element.
% The distance between $CHD_1$ and $CHD_2$ is shown in the figure.

% [figure]

% The distance increases along frequency linearly and positions exponentially. This result indicates that the distance is more sensitive to the duplicates of the larger element, which is a suitable property for structural embedding.

\subsection{Estimating structural features on multiple hops}
In addition to the edge size information discussed previously, it is also necessary to consider the connection patterns of neighbors to assess structural features.
To compare the structural similarity of two given nodes at $k-$hop, we collect the $CHD$ of the neighbors at $k-$hop ($NCHD^k$).
Given a node $u$ and a hop number $k$, $NCHD^k(u) = \{CHD_j | j \in N^k(u)\}$.
$N^k(u)$ indicates the $k-$hop neighbors from node $u$.

We then calculate the distance ($D^{'k}(u,v)$) between two nodes' $NCHD^k$ by applying the dynamic time wrapping function, as the length of $NCHD^k$ (the number of $k-$hop neighbors) can vary. 
\begin{equation}
    D^{'k}(u,v)=DTW\big(D_0;\enspace NCHD^k(u),\enspace NCHD^k(v)\big).
    \label{equ:dis_k_ncmpd}
\end{equation}
Note that $NCHD^k$ is a list of $CHD$; we further compress $NCHD^k$ by its frequency to form the tuple list $CNCHD^k = [(CHD_1^k,f_1^k),(CHD_2^k,f_2^k),\cdots]$, as CHD to HD, to reduce the computational cost.

\begin{equation}
    CNCMPD(CHD^k_i, CHD^k_j) = max(f_i^k, f_j^k) \times DTW(D^0;\enspace CHD_i,\enspace CHD_j).
    \label{equ:cncmpd}
\end{equation}
\begin{equation}
    D^k(u,v)=DTW\big(CNCMPD;\enspace CNCHD^k(u),\enspace CNCHD^k(v)\big).
    \label{equ:dis_k_cncmpd}
\end{equation}

\subsection{Creating a multilayer graph to generate structural embeddings}
We apply the framework proposed in struc2vec to generate structural embeddings, and calculate the multi-scale distances for each node pair in different hops.
\begin{equation}
    dis^{k}(u, v) = dis^{k-1}(u, v) + DTW\Big(D;\enspace sd\big(N^k(u)\big),\enspace sd\big(N^k(v)\big)\Big),
    \label{formula:s2v_distance}
\end{equation}
where $dis_{-1}=0$, $sd(\cdot)$ is the sorted degree. $D$ is the distance function used in DTW, and is equation \ref{equ:dis_0_cmpd} for $k=0$ and equation \ref{equ:dis_k_cncmpd} for $k>0$.

We create a $k-$layers-weighted graph so that each layer contains all the nodes and represents a specific hop.
The following equation then calculates the weight between nodes u and v within the $k$th layer.
\begin{equation}
    w^k(u, v) = e^{-dis^k(u,v)}.
    \label{formula:s2v_w_inside}
\end{equation}
For node $u$ in the $k$th layer, the weights moving up to the $k+1$th layer and down to the $k-1$th layer are calculated from the following equations.
The weight between two nodes in the same layer represents the structural similarity, while those connecting the same nodes between layers represent the overall quantity of similar nodes within a specified layer.
\begin{equation}
    w(u^k, u^{k+1}) = \log{\big(\Gamma^k{(u)} + e\big)},  
    \label{formula:s2v_w_up}
\end{equation}
\begin{equation}
    w(u^k, u^{k-1}) = 1.
    \label{formula:s2v_w_down}
\end{equation}
$\Gamma^k{(u)}$ represents the count of edges connected to node $u$ that possess weights greater than the average weight of all edges in the $k$th layer.
% The edge's weight between two nodes represents the structural similarity and is calculated as equation XX.
% Edges connect the same nodes between layers, and the weights are calculated by equation XX.

We generate the walk sequences by applying a biased random walk on the $k-$layers-weighted graph. If there are similar nodes to the current node in the current layer, the walker tends to move to these nodes within the same layer; otherwise, the walker moves up or down to find similar nodes in different hops. 
The probabilities are calculated from the following equations.
\begin{align}
    & p^k(u,v)=q\frac{e^{-dis^k(u,v)}}{\sum_{v\in{V}, v\ne u}{e^{-dis^k(u,v)}}}, \\
    & p^k(u^k, u^{k+1}) = (1-q)\frac{w(u^k, u^{k+1})}{w(u^k, u^{k+1}) + w(u^k, u^{k-1})} \label{formula:s2v_p_up}  ,  \\
    & p^k(u^k, u^{k-1}) = (1-q)\big(1 - p^k(u^k, u^{k+1})\big) \label{formula:s2v_p_down}.
\end{align}
$q$ is the probability of staying in the current layer.
In this way, walk sequences containing similar nodes in multiple scales are expected to be generated. 

Lastly, we apply Skip-Gram \cite{mikolov2013efficient} on these sequences to generate the embeddings of nodes under the assumption that similar nodes share similar embeddings.

\section{Experiments}
We then conducted various experiments on toy and real networks to confirm the performance of HyperS2V.
\subsection{Visualization of embeddings on toy networks}
\subsubsection{Dataset}

\begin{figure}
    \centering
    \includegraphics[scale=0.20]{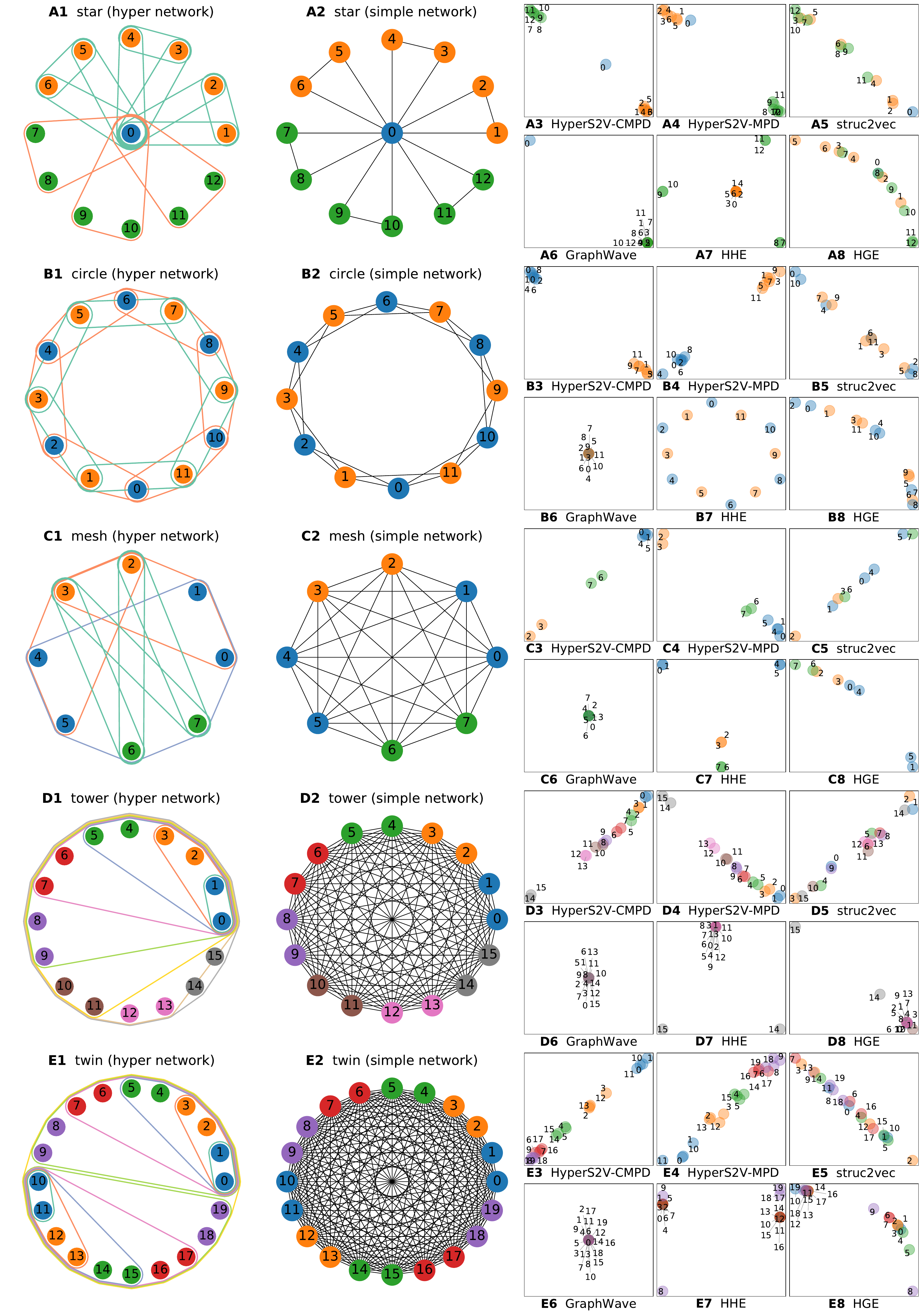}
    \caption{Visualization results}
    \label{fig:exp_toy_all}
\end{figure}

We created five toy hyper networks, as shown in Figure \ref{fig:exp_toy_all} A1-E1: star, circle, mesh, tower, and twin. Figure \ref{fig:exp_toy_all} A2-E2 shows the corresponding simple networks yielded by clique expansion; namely, hyperedges are converted into cliques. 
% Hyperedges are colored by size. 
We color nodes by their structural features in hyper networks, and copy the color over to the simple networks.
Thus, nodes with the same color share the same structure in hyper networks but may not be in simple networks due to the information loss from clique expansion.
\subsubsection{Comparison methods and hyperparameters}
We selected embedding methods for hyper networks (HHE, HGE) and simple networks (GraphWave, struc2vec) to perform our comparisons with HyperS2V. We use hyper networks as inputs for HHE, HGE, and the proposed method, and simple networks for GraphWave and struc2vec.
We tested two variations of HyperS2V: HyperS2V-CMPD (distances calculated from CHD and CNCHD using equations \ref{equ:dis_0_cmpd} and \ref{equ:dis_k_cncmpd}), and HyperS2V-MPD (distances calculated without any collapsing).
The hyperparameters of HyperS2V are set as follows: 100 walks per node, walk length of 80, window size of 5, and maximum layer of 5. The hyperparameters of comparison methods are kept as default.
\subsubsection{Results}

We generated two-dimensional embeddings from all methods and plotted the results in Figure \ref{fig:exp_toy_all} A3-E8.
Since nodes are colored by structural features, it is expected that the points with the same color are clustered together in plots. HyperS2V was able to form clusters by colors in all cases, indicating that the embeddings preserved the high-order structural information. HyperS2V-CMPD performed even better than HyperS2V-MPD, placing clusters evenly in the star and mesh plots. All the comparison methods failed to cluster nodes by colors across all networks. We suggest that the proximity-based nature of HHE and HGE resulted in these failures, while the information loss of the simple networks induced the results of GraphWave and struc2vec.

\subsection{Hyperedge size prediction on real networks}

\subsubsection{Dataset}

\begin{table}[t]
    \caption{Statistics of hyper networks}
    \centering
    \begin{tabular}{l|r|r|r|r|r|r} \hline
         dataset & \#node & \#edge & aspect ratio & density & max degree & max edge  \\ \hline
         Zoo & 101 & 43 & 2.35 & 0.3953 & 17 & 93  \\ \hline
         House & 1,290 & 341 & 3.78 & 0.0269 & 44 & 81  \\ \hline
         Cora-cited & 2,094 & 946 & 2.21 & 0.0024 & 5 & 166  \\ \hline
         Cora-citing & 1,434 & 1,579 & 0.91 & 0.0021 & 145 & 5  \\ \hline
         GS-hyper1 & 3,026 & 5,015 & 0.60 & 0.0009 & 414 & 44  \\ \hline
         GS-hyper2 & 8,019 & 19,569 & 0.41 & 0.0003 & 665 & 42  \\ \hline
         GS-hyper3 & 5,844 & 14,023 & 0.42 & 0.0005 & 825 & 28  \\ \hline
         GS-hyper4 & 1,054 & 1,655 & 0.64 & 0.0025 & 171 & 26  \\ \hline
         GS-hyper5 & 381 & 964 & 0.40 & 0.0064 & 423 & 19  \\ \hline
         GS-hyper6 & 722 & 1,737 & 0.42 & 0.0038 & 464 & 23  \\ \hline
         GS-hyper7 & 9,293 & 17,160 & 0.54 & 0.0003 & 604 & 44  \\ \hline
         GS-hyper8 & 6,113 & 21,855 & 0.28 & 0.0004 & 793 & 35  \\ \hline
         GS-hyper9 & 5,742 & 11,926 & 0.48 & 0.0005 & 780 & 44  \\ \hline
         GS-hyper10 & 4,147 & 7,996 & 0.52 & 0.0007 & 495 & 28  \\ \hline
         Lesmis & 77 & 157 & 0.49 & 0.0371 & 39 & 9  \\ \hline
         GS-hyper All & 337,450 & 2,632,829 & 0.13 & 0.00001 & 1,387 & 61  \\ \hline
    \end{tabular}
    \label{tab:stat_hyper}
\end{table}
We conducted hyperedges' size prediction experiments based on five real networks grounded in life and social sciences, to citation and co-authorship. These are the Zoo network\footnote{https://archive.ics.uci.edu/dataset/111/zoo} \cite{misc_zoo_111}, House network\footnote{http://web.mit.edu/17.251/www/data\_page.html} \cite{chodrow2021house-committe,chodrow2021hypergraph}, Cora-cited network, Cora-citing network, and GS-hyper subnetworks.
The Zoo hyper network is created from the Zoo dataset based on life sciences. Nodes are animals, and hyperedges are formed among the animals sharing the same values for a specified attribute. 
In the House hyper network, each hyperedge represents a committee within a session of Congress, with each node symbolizing a House member.
Cora-cited and Cora-citing are hyper networks created from the Cora dataset\footnote{https://relational.fit.cvut.cz/dataset/CORA} \cite{sen:aimag08}. They are citation networks in which the nodes are papers. Papers cited by the same article are encircled by a hyperedge in the Cora-cited network.
The GS-hyper network is a huge co-author network that we created based on the raw data of the GS dataset by Chen et al. \cite{Chen_BigScholar17}.
Chen et al. scraped co-authorship data from Google Scholar and provided a co-authorship (simple) network in which nodes are researchers and edges between two researchers are created if they co-authored any papers.
We created the GS-hyper network where the hyperedges represent exact papers. Since the original GS-hyper network (``GS-hyper All" in Table \ref{tab:stat_hyper}) is too big, we randomly selected ten nodes and collected their 4-hops neighbors to form ten subnetworks (``GS-hyper 1-10" in Table \ref{tab:stat_hyper}) with the proper size. 
% The statistics are shown in Table XX, and the node degree and edge size distributions are plotted in Figure XX.

\subsubsection{Comparison methods, hyperparameters, and experiment setting}
% For the experiments on real networks from now on, we use HyperS2V to indicate the distances of CHD and CNCHD are calculated by equation \ref{equ:dis_0_cmpd} and \ref{equ:dis_k_cncmpd} (same as HyperS2V-CMPD in the previous experiment.)
Regarding experiments involving real networks, we employ HyperS2V to denote that the distances of CHD and CNCHD are computed using equations \ref{equ:dis_0_cmpd} and \ref{equ:dis_k_cncmpd} respectively. This choice aligns with the approach adopted in the earlier HyperS2V-CMPD experiment, owing to its enhanced performance and reduced computational overhead.
We used the comparison methods used in the previous experiment except for HGE; this was due to HGE failing to converge in these datasets.
The hyperparameters were kept the same as in the previous experiment.
We embedded these datasets into a 64-dimensional space and created the hyperedges embeddings by taking the average of the corresponding nodes' embeddings.
We then fed 80\% of the hyperedges embeddings into a ridge regression model to predict the edges' sizes. We confirmed the average root-mean-square error (RMSE) as the prediction result on the rest data.

\subsubsection{Results}

\begin{table}[t]
    \caption{Results on hyperedge size prediction}
    \centering
    \begin{tabular}{l|c|c|c|c} \hline
         dataset & HyperS2V & struc2vec & HHE & GraphWave  \\ \hline
         Zoo & \bf{24.017}  & 27.524 & 27.246 & 27.501  \\ \hline
         House & \bf{8.439}  & 15.558 & 21.240 & 21.396  \\ \hline
         Cora-citing & \bf{0.822}  & 0.955 & 1.022 & 1.024  \\ \hline
         Cora-cited & \bf{6.561}  & 7.200 & 7.141 & 7.147  \\ \hline
         GS-hyper1 & \bf{1.284}  & 1.713 & 1.905 & 1.935  \\ \hline
         GS-hyper2 & \bf{0.901}  & 1.139 & 1.221 & 1.234  \\ \hline
         GS-hyper3 & \bf{1.409}  & 1.966 & 2.077 & 2.095  \\ \hline
         GS-hyper4 & \bf{1.335}  & 1.534 & 1.761 & 1.771  \\ \hline
         GS-hyper5 & \bf{1.028}  & 1.139 & 1.186 & 1.228  \\ \hline
         GS-hyper6 & \bf{0.843}  & 1.180 & 1.467 & 1.503  \\ \hline
         GS-hyper7 & \bf{1.215}  & 1.539 & 1.624 & 1.639  \\ \hline
         GS-hyper8 & \bf{0.908}  & 1.110 & 1.201 & 1.214  \\ \hline
         GS-hyper9 & \bf{1.366}  & 1.774 & 1.824 & 1.838  \\ \hline
         GS-hyper10 & \bf{1.267}  & 1.800 & 1.958 & 1.988  \\ \hline
    \end{tabular}
    \label{tab:exp_hyperedge_size}
\end{table}
The results of all methods are shown in Table \ref{tab:exp_hyperedge_size}; the smaller the RMSE, the better the performance. 
HyperS2V performed the best in all datasets, indicating that the embeddings preserved the most structural information of hyperedges compared with the other methods. Struc2vec achieved the second-best performance even though it is not proposed for hyper networks; we found that struc2vec had indirectly preserved the hyperedge structure from the cliques.

\subsection{Hyperedge prediction}

\subsubsection{Dataset and experiment setting}

We conducted hyperedge prediction on the same real networks as in the previous experiment.
We randomly sampled non-existent hyperedges with the same numbers of existing ones on each hyperedge size and generated the hyperedges' embeddings by averaging the corresponding nodes' embeddings.
We trained logistic regression prediction models by 80\% of the data and evaluated the AUC score of the rest. Thus, the higher the AUC, the better.
\subsubsection{Results}
\begin{table}[t]
    \caption{Results on hyperedge prediction}
    \centering
    \begin{tabular}{l|r|r|r|r} \hline
         dataset & HyperH2V & struc2vec & HHE & GraphWave  \\ \hline
         Zoo & 0.493 & 0.351 & 0.406 & \bf{0.500}   \\ \hline
         House & 0.400 & 0.355 & 0.442 & \bf{0.486}   \\ \hline
         Cora-cited & \bf{0.590}  & 0.479 & 0.419 & 0.513  \\ \hline
         Cora-citing & \bf{0.637}  & 0.591 & 0.470 & 0.612  \\ \hline
         GS-hyper1 & \bf{0.839}  & 0.703 & 0.554 & 0.698  \\ \hline
         GS-hyper2 & \bf{0.829}  & 0.727 & 0.574 & 0.710  \\ \hline
         GS-hyper3 & \bf{0.846}  & 0.701 & 0.601 & 0.693  \\ \hline
         GS-hyper4 & \bf{0.821}  & 0.739 & 0.551 & 0.681  \\ \hline
         GS-hyper5 & \bf{0.830}  & 0.782 & 0.624 & 0.749  \\ \hline
         GS-hyper6 & \bf{0.847}  & 0.787 & 0.588 & 0.679  \\ \hline
         GS-hyper7 & \bf{0.815}  & 0.704 & 0.516 & 0.708  \\ \hline
         GS-hyper8 & \bf{0.846}  & 0.761 & 0.586 & 0.721  \\ \hline
         GS-hyper9 & \bf{0.839}  & 0.702 & 0.586 & 0.672  \\ \hline
         GS-hyper10 & \bf{0.851}  & 0.709 & 0.578 & 0.704  \\ \hline
 
    \end{tabular}
    \label{tab:exp_edge}
\end{table}
The results are shown in Table \ref{tab:exp_edge}.
HyperS2V outperformed our comparison methods in most datasets (except Zoo and House), demonstrating considerable improvement. This demonstrates that HyperS2V has the capability of preserving higher-order structural properties. 
GraphWave performed the second-best in citation networks while struc2vec performed the same in co-author networks, indicating that even structural embedding methods for simple networks could partially grasp high-order structural properties.
In the case of Zoo and House datasets, even the best-performing method could only achieve an AUC equal to or smaller than 0.5. We suggest the reason might be that these datasets contain hyperedges with large sizes in contrast to the number of nodes, therefore it is difficult to sample non-existent hyperedges to the same size. A similar observation exists in the Cora-cited network, where the prediction result is also relatively low.
A secondary reason is also that the hyperedges are formed through a broader meaning instead of interactions, such as the node attributes in the Zoo data, Congress committees in the House data, and co-citation in Cora-cited networks.
% the semantic meaning of the hyperedges in these two datasets, i.e., aggregations of features but not direct interactions between nodes.
Furthermore, hyperedges represent different semantic meanings even in one hyper network, such as in Zoo.
% , hyperedges could exist among animals with two legs and animals with tails
These properties lead to the mismatch of the semantic meaning of the hyperedges and the hyperedge prediction task setting; in short, it is not appropriate to predict one attribute by using others.

\subsection{Case study on Les Misérables}
\subsubsection{Dataset and experiment setting}
We conducted a case study on the Les Misérables dataset\footnote{http://ftp.cs.stanford.edu/pub/sgb/jean.dat} \cite{knuth1993stanford}, based on the English version of Hugo Victor's Les Misérables, a French historical novel that follows the life and struggles of Jean Valjean and several other major characters.
Nodes are characters, and hyperedges are the interactions, i.e., a group of characters is in one hyperedge if they appear in one scene.
We eliminated the duplicated hyperedges and obtained the largest component of the Lesmis hyper network, with the statistics shown in Table \ref{tab:stat_hyper}.
We applied embedding methods on lesmis to gain 2-dimension vectors for each node. We then clustered the embeddings by k-means clustering into six clusters because six is a proper number for all methods.
% by leveraging the information kept and group numbers.

\subsubsection{Results}

\begin{figure}[t]
    \centering
    \includegraphics[scale=0.33]{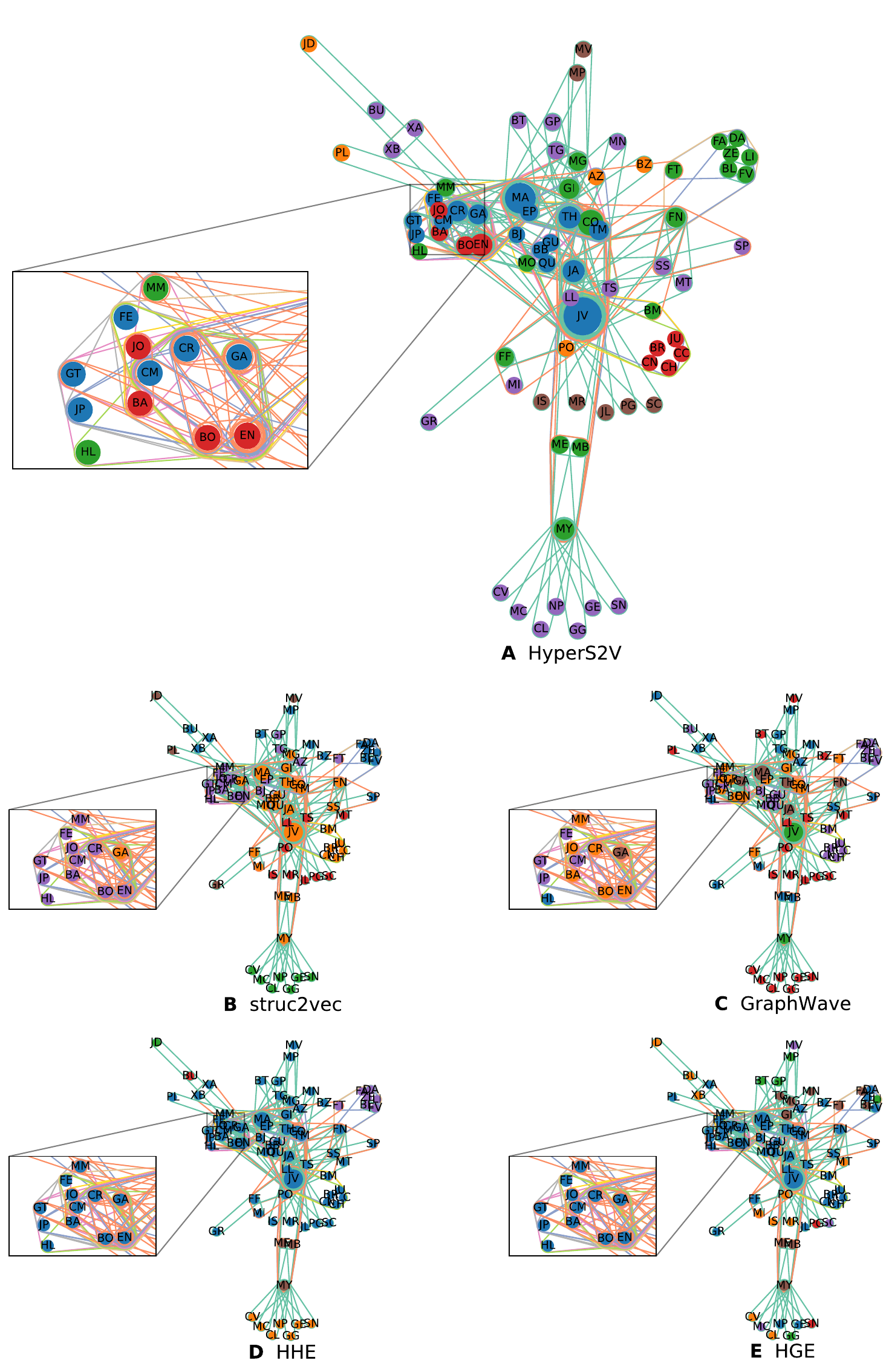}
    \caption{Clustering results on Les Misérables}
    \label{fig:les_mis}
\end{figure}
The results in Figure \ref{fig:les_mis} suggest that HyperS2V outperformed the other methods. HyperS2V identified appropriate clusters, and also identified nuances within clusters as well. Graphwave performed modestly well, although a number of clusters were inexplicably linked. For example, the group of lovers (FA, DA, ZE, LI, BL, FV), the participants in the trial of CH (CN, BR, JU, CC, CH), minor members of the Patron-Minette gang (BJ, MO, BB, GU, QU), and certain members of the Friends of the ABC (GT, JP, FE, and CM) are grouped in the same cluster, despite no obvious interactions between these groups. Struc2vec also performed reasonably, though it failed to capture nuances within certain groups that HyperS2V and Graphwave identified. It was also generally overly broad. Struc2vec was also poorly balanced, as it differentiated minor fringe characters (JD/BU, MV/MP as examples) at the expense of grouping other minor characters with major characters such as JV and JA. HHE and HGE performed the worst, being too broad in grouping major and minor characters together, and differentiating individuals with similar interactions. For example, in HHE, many characters grouped in the same cluster as the main character, Jean Valjean (JV), can be characterised by simply having interacted directly with him at some point in the novel. It does not distinguish social groups, such as the previously mentioned Patron-Minette criminal gang led by TH and TM, the friends of the ABC group who lead the failed uprising towards the end of the series (EN, BO, CR, CM, BA, JO, etc.), or those involved in the trial. HGE performs similarly poorly, and inexplicably differentiates CV, MC, etc. at the bottom despite all these individuals playing minor roles, with interaction only with the bishop MY.

Given the superior performance of HyperS2V, we examine the results of this method in more detail. HyperS2V distinguishes the main characters and surrounding clusters better than our other methods. As part of the main story, Jean Valjean (JV), the novels primary protagonist is pursued by Javert (JA), the police officer, throughout the course of the novel. While both characters encounter almost all of the other clusters at some point, it is arguably the interactions between the two, the Thernadiers (TH, TM) who lead the Patron-Minette gang, and Marius (MA) who co-leads the Friends of the ABC group that are the most significant. Major events include Valjean freeing Cosette (CO), daughter of Fantine (FN), from her indentured servitude of the Thernadiers, and saving Marius' life. Valjean and Javert both also take part in the failed uprising of the Friends of the ABC, though Javert does so as a spy.

We use the aforementioned Friends of the ABC group as an example of where HyperS2V picks up nuances within the clusters to identify different roles that individuals played within a group. In addition to the aforementioned individuals of the ABC, there is also Gavroche (GA), a street urchin who is the son of the Thernadiers. While not a member of the ABC, he fights and dies alongside them in the uprising led by the ABC. Comparing to struc2vec's grouping of the ABC members for simplicity, we observe that the interactions of members within and close to the ABC group result in them being grouped together. Monsieur Mabeuf (MM) and Madame Hucheloup (HL) are not directly part of the ABC group, though they are involved in the uprising that occurs in Volume Four in some way, and is considered a climactic point in the novel. We observe that struc2vec was not necessarily incorrect in grouping these individuals with the rest of the ABC members. It is however, overly simplistic. Madame Hucheloup (HL) only owns the inn that the ABC members use, and does not take part in the uprising herself. Similarly, Monsieur Mabeuf (MM) is a prefect of the church who falls into destitution and joins the uprising when he feels he has nothing left to lose. HyperS2V was able to pick up these subtleties within the group, differentiating these two characters. Graphwave also differentiates HL, however MM is grouped with other members of the Friends of the ABC.

HyperS2V fell short in only one regard; some groups with no connection were placed in the same cluster. However, this was observed across all methods that relied on structural characteristics (i.e. GraphWave and struc2vec). In structural based methods, nodes can be placed in the same cluster if they interact with surrounding nodes in a similar way. In the case of HyperS2V, this suggests that the bishop MY and his cluster ME and MB along with GI, MG, and MO interact with their surrounding characters in similar ways, despite no connection in regards to the story. This characteristic may also have allowed HyperS2V and Graphwave to differentiate HL and MM, as previously indicated. Through this analysis, we demonstrate that HyperS2V is able to identify key roles and clusters. Further, the ability to identify structurally similar yet unrelated roles is an opportunity to further our understanding of the network and how it relates to the story.

\section{Conclusion}
Both hyper networks and structural information exhibit superior expressive capabilities; their combination holds promise for advancing the realms of network science and machine learning. 
We proposed HyperS2V, a novel structural embedding method for nodes in hyper networks. We pioneeringly proposed HD (hyper-degree) and NCHD (neighbors' collapsed hyper-degree) to represent the 0- and further hops structure of a node in hyper networks, and MPD (magnitude-position distance) function to calculate the similarity between HDs (NCHDs) by leveraging the magnitude and position of the elements inside. Moreover, we conducted a series of experiments to demonstrate the high performance on interpretability from the visualizations of embeddings from toy networks and its high adaptability to downstream tasks such as hyperedge (dimension) prediction. Furthermore, we employ HyperS2V in a case study of the Les Misérables dataset, underscoring its practical utility.
% Besides, we directly constructed GS-hyper, a new co-authorship dataset in the form of hyper networks, from the raw data scraped from Google Scholar. It contains attribute labels of both nodes and hyperedges. GS-hyper is expected to be widely used in hyper-network-based analyses in the science of science field and other machine learning fields.

There are several drawbacks of HyperS2V. The first and most significant one is the need for more fundamental research on hyper networks in the real world. Hyper networks are more flexible to the size of hyperedges, from which different types can be derived such as hyper networks with a small number of nodes and lots of hyperedges, or a large number of nodes and few hyperedges. This flexibility makes it difficult to define a universe distance function. Although we proposed MPD as our current best method for the lowest information loss when integrating hyperedge sizes' magnitude and position, MPD could be considered inappropriate for particular hyper networks, such as the Zoo network. By further acknowledging the nature of hyper networks in the real world, we hope to encourage further development of research into structural features of hyper networks.
Another drawback is the scalability of HyperS2V. NCHD, representing a structure of one node in one or further hops, is a list of list-of-scalers, and could preserve the most structural information. However, it significantly increases the computational cost. Consequently, the scalability of HyperS2V becomes a weakness. To address this, we contemplate approximating similarity calculation functions and exploring faster programming languages than Python for implementation.

\section*{Acknowledgment}
This work was supported by JST SPRING, Grant Number JPMJSP2108.
\section*{Availability of data and materials}The implementation codes and data are available on GitHub (\url{https://github.com/liushu2019/HyperS2V.git}).
\section*{Author Contributions}S.L. conceived and designed the research, developed the methodology, performed the experiments, analyzed the results, and wrote the manuscript; C.L. partially analyzed the results and wrote the manuscript; S.L. and F.T. obtained funding; F.T. discussed the result and provided reviews. All authors read and approved the final version of the paper.
\section*{Competing interests}The authors declare no competing interests.

\bibliographystyle{unsrt}
\bibliography{liu_bib_main}
\end{document}